\documentclass{PoS}
\usepackage[utf8]{inputenc}
\usepackage{xcolor}
\usepackage{lineno}
\usepackage{graphicx}
\usepackage{subcaption}
\usepackage[section]{placeins}
\usepackage[normalem]{ulem}
\usepackage{url}
\usepackage{wrapfig}
\title{Search for Dark Matter Annihilation to Neutrinos from the Sun}

\ShortTitle{IceCube Solar WIMP Annihilation}

\author{
The IceCube Collaboration\footnote{For collaboration list, see PoS(ICRC2019) 1177.}\\
{\itshape \href{http://icecube.wisc.edu/collaboration/authors/icrc19_icecube}{http://icecube.wisc.edu/collaboration/authors/icrc19\_icecube}}\\
E-mail: \email{caad@mit.edu, akheirandish@icecube.wisc.edu, jeffrey.lazar@icecube.wisc.edu, qliu@icecube.wisc.edu}
}

\abstract{
Weakly interacting massive particles (WIMPs) can be gravitationally captured by the Sun and trapped in its core. The annihilation of those WIMPs into Standard Model particles produces a spectrum of neutrinos whose energy distribution is related to the dark matter mass. In this work, we present the theoretical framework for relating an observed neutrino flux to the WIMP-nucleon cross section and summarize a previous solar WIMP search carried out by IceCube. We then outline an ongoing updated solar WIMP search, focusing on improvements over the previous search.\\

\vspace{4mm}
{\bfseries Corresponding authors:}
Carlos A. Arg{\"u}elles$^{1}$, Ali Kheirandish$^{2}$, \speaker{Jeffrey Lazar}$^{2}$, Qinrui Liu$^{2}$\\
{$^{1}$ \itshape Dept. of Physics, Massachusetts Institute of Technology, Cambridge, MA 02139, USA}\\
{$^{2}$ \itshape Wisconsin IceCube Particle Astrophysics Center and Dept. of Physics, University of Wisconsin, Madison, WI 53706, USA}
}

\FullConference{36th International Cosmic Ray Conference -ICRC2019-\\
		July 24th - August 1st, 2019\\
		Madison, WI, U.S.A.}

\begin{document}
\section{Introduction}\label{sec:info}
Astrophysical and cosmological observations provide strong evidence for the existence of dark matter (DM). However, the nature of DM remains unknown.
One class of DM candidates is weakly interacting massive particles (WIMPs), which are expected to have masses from a few GeV to a few TeV; see \cite{Bertone:2004pz} for a comprehensive review.
If WIMPs make up DM, they could scatter off nuclei inside large celestial bodies (which in this case refers to the Sun), losing energy in the process and possibly becoming gravitationally bounded. 
Then, the WIMPs could undergo additional scatterings, and fall to the center of the Sun \cite{1985ApJ...296..679P, PhysRevD.34.2206, Ritz:1987mh, SREDNICKI1987804}. 
As WIMPs continue to be captured, an excess accumulates at the core of the Sun. 
The large density of WIMPs allows them to annihilate effectively to Standard Model (SM) particles, creating neutrinos directly or through subsequent interactions.
For WIMPs with mass above a few GeV, we expect the capture and annihilation processes to equilibrate, \textit{i.e.}, after long enough time, the annihilation rate is half of the capture rate. \cite{Jungman:1995df}. 
Since the capture rate depends only on the WIMP-nucleon cross section, $\sigma_{\chi n}$ \cite{Gould:1991hx}, constraining the neutrino flux constrains the WIMP-nucleon cross section.

To detect this flux, the IceCube Neutrino Observatory, located at the South Pole, can look for an excess of $\nu_{\mu}$ and $\bar{\nu}_{\mu}$ from the direction of the Sun. 
The region in which the neutrinos are produced is small enough that this amounts to a point source analysis.
Since for neutrino energies greater than 1 TeV the mean free path is less than the radius of the Sun, we expect the signal to appear in the energy range of a few GeV to $\sim$ 1 TeV.
In this energy range, the only intrinsic background from the Sun are neutrinos produced in interactions between cosmic rays and the solar atmosphere. 
Additionally, we expect a background from cosmic ray interactions in the Earth's atmosphere.
The uncertainty in the intrinsic background is at the level of 30$\%$ \cite{Ingelman:1996mj, Arguelles:2017eao, Ng:2017aur, Edsjo:2017kjk}, thus an observation of an excess above terrestrial and solar atmospheric backgrounds would be compelling evidence for new physics. 
%This is unique with respect to other astrophysical scenarios such as galactic halo searches because the background is well-understood.
This contrasts with other indirect searches, such as multimessenger detections from the galactic halo \cite{Aartsen:2014hva, Abeysekara:2017jxs}, where backgrounds are weakly constrained.

\section{Theoretical Framework and Previous IceCube Solar WIMP Search}
When considering the number of WIMPs in the Sun, the effect of WIMP evaporation can be ignored when the mass is higher than a few GeV \cite{Jungman:1995df}.
In this case, the number of WIMPs inside the Sun is approximately determined by the capture and the annihilation rates.
Since IceCube is most sensitive to dark matter masses above approximately 100 GeV, evaporation can be neglected in IceCube solar WIMP searches. In this regime, the number of WIMPs, $N_{\rm{DM}}$, is given by:
\begin{equation}\label{eq:Nwimp}
\frac{dN_{\rm{DM}}}{dt}= C_{\odot}-A_{\odot}N_{\rm{DM}}^2, 
\end{equation}
where $C_{\odot}$ is the capture rate and $A_{\odot}$ is the product of the annihilation cross section and the relative WIMP velocity divided by the effective volume of the core of the Sun.
For timescales on the order of the lifetime of the Sun, $t_\odot$, the distribution of WIMPs has equilibriated.
In equilibrium, the number of WIMPs in the Sun is constant. 
This implies that the capture rate and annihilation rate, $\Gamma_{\rm{ann}}$, are related by 
\begin{equation}\label{eq:capture_ann}
\Gamma_{\rm{ann}} = \frac{1}{2}A_\odot N_\mathrm{DM}(t_\odot)^2\approx\frac{C_\odot}{2} \propto \sigma_{\chi N}.
\end{equation}
 
The capture rate could be due to spin-independent or spin-dependent WIMP scattering off nuclei and is calculated in \cite{Jungman:1995df, Gould:1991hx}.
The calculation of the capture rate depends not only on the solar density but also on the composition of the Sun.
For spin-independent interactions, WIMPs interact coherently and the cross section grows with the square of the target atomic number.
This implies a preference for WIMPs to scatter off heavy nuclei.
On the other hand, spin-dependent scattering is enhanced by nuclei which have a nonzero net spin, making hydrogen a good target.
Since hydrogen is the most abundant element in the Sun, the Sun is expected to provide a good constraint on the spin-dependent WIMP-nucleon cross section.

IceCube previously carried out a solar WIMP search using data from three years of livetime in the full 86-string configuration \cite{Aartsen:2016zhm}.
The main sources of systematic uncertainty in this analysis are due to uncertainties in the neutrino-nucleon cross section, the neutrino oscillation parameters, muon propagation in the ice, absolute DOM efficiency, and photon propagation in the ice.
While this analysis did not find a significant excess, IceCube's current limits on the spin-dependent WIMP-nucleon cross section are the most stringent for WIMPs above 80 GeV annihilating to $b\bar{b}$, $W^{+}W^{-}$, and $\tau^{+}\tau^{-}$.
\section{Analysis}
In our analysis, we look to improve upon the previous search and maximize our chances of detecting dark matter.
To do this, we first updated the solar model used in calculating the capture rate to the latest version in \cite{Vinyoles:2016djt}.
Next, we utilize a new high-purity event selection with the most up-to-date systematic treatment and updated physical parameters.
Furthermore, we incorporate the additional neutrino background from cosmic ray interactions in the solar atmosphere, which could mimic signal.
%, and include systematic uncertainties that arise from the choice of signal generation software.
Lastly, we study different signal generation packages to understand how these contribute to our systematic uncertainties.
\subsection{Neutrino Flux Study}
After the annihilation of WIMPs in the center of the Sun, SM particles are produced through different annihilation channels. 
Neutrinos are generated after hadronization and decays of those SM particles. We focus on three benchmark annihilation channels: $b\bar{b}$, $W^+W^-$, and $\tau^+\tau^-$, because they represent both soft and hard neutrino spectra.

Expected neutrino generation from WIMP annihilation has been computed in WimpSim \cite{Blennow:2007tw}, Cirelli \textit{et al.} \cite{Cirelli:2005gh}, and more recent PPPC \cite{Baratella:2013fya}, and it can also be computed using Monte Carlo (MC) generators, such as PYTHIA \cite{Sjostrand:2006za, Sjostrand:2014zea}, directly. Here, we generate neutrinos using PYTHIA8240, which is the latest PYTHIA version, and compare that with other computations mentioned above.
%\pagebreak
\subsubsection{Neutrino Production}
Because SM particles are produced in the solar core, which has densities of $\mathcal{O}(10^2) \mathrm{gr/cm}^3$~\cite{Vinyoles:2016djt}, they experience interactions before decaying into, among other things, neutrinos. Interactions are not included in MC generators designed for colliders like PYTHIA, so they have to be separately implemented. In the case of heavy quarks, the top quark decays before it interacts because of its extremely short lifetime, whereas bottom and charm quarks can hadronize. Hadrons composed of bottom or charm quarks have interaction lengths comparable to decay lengths in the Sun's center, both of which are on the order of millimeters, and therefore energy losses have to be considered. We use the average energy loss estimated in \cite{Ritz:1987mh}.

\begin{wrapfigure}{r}{0.5\textwidth}
%\begin{figure}
\centering
\includegraphics[width=0.8\linewidth]{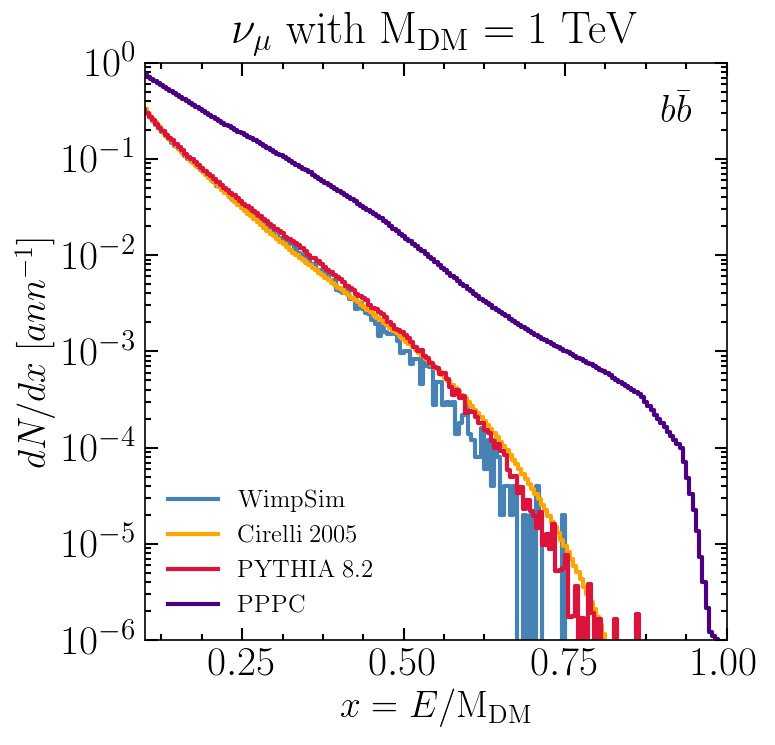}
\includegraphics[width=0.8\linewidth]{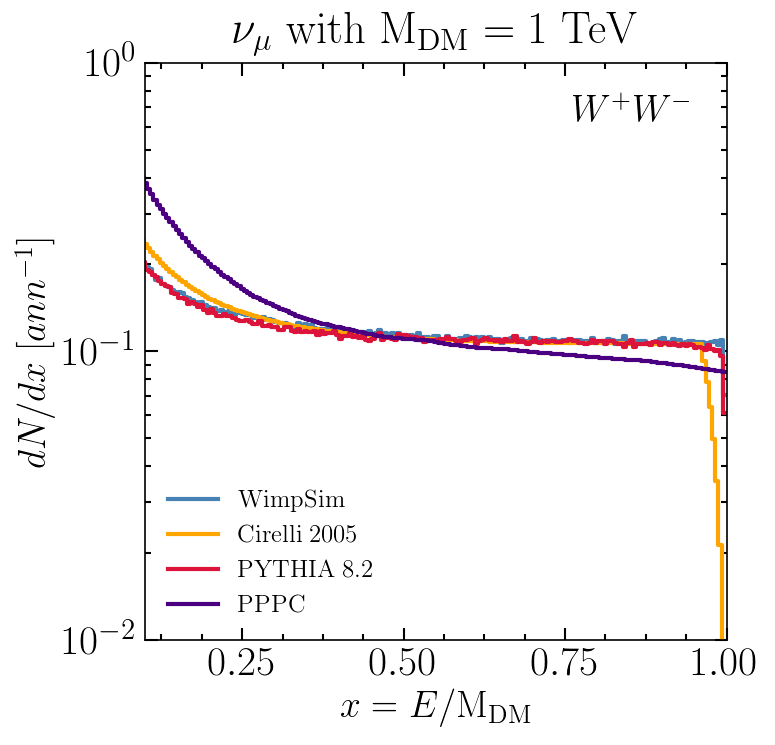}
\includegraphics[width=0.8\linewidth]{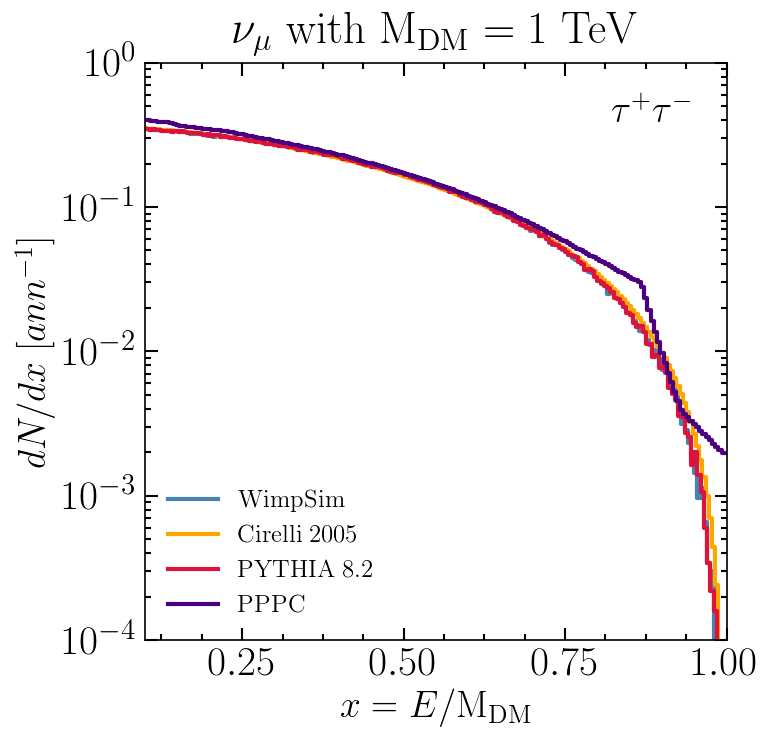}
 %\vspace{1em}
\caption{$\nu_\mu$ yields per annihilation in the Sun core from different neutrino production calculations for three WIMP annihilation channels as a function of $x=E/\mathrm{M_{DM}}$ when $\mathrm{M_{DM}}=1000\,\mathrm{GeV}$.}
\label{fig:nuproduction}
%\end{figure}
\end{wrapfigure}
WimpSim and Cirelli \textit{et al.} use the same approach, while PPPC uses a distribution of energy transfer. Hadrons composed of only light quarks, such as $\pi$ and $K$, have a relatively long lifetimes, which result in much longer decay lengths compared with their interaction lengths. Therefore, they rapidly lose energy before decaying. For charged leptons, the stopping time in plasma has to be compared with their lifetimes. For the Sun, $\tau$ would only have negligible energy loss while most $\mu$ would stop completely before decaying. Light hadrons and leptons that experience numerous interactions can be ignored in IceCube since they only produce neutrinos below GeV. When the mass of the WIMP reaches the electroweak (EW) scale, there would be EW splitting, which affects neutrino production. This EW correction is considered in PPPC but not in our computation. By using PYTHIA8240 \cite{Sjostrand:2014zea}, we also incorporate the latest particle data table into neutrino generation. In Fig. \ref{fig:nuproduction}, $\nu_\mu$ yields per WIMP annihilation with calculations discussed above are compared. %As shown in Fig. \ref{fig:nuproduction}, WimpSim, Cirelli \textit{et al.} and what we generated approximately agree while PPPC is pretty different. The difference can be cause by the treatment of bottom/charm hadron interactions and EW correction. Understanding of uncertainties from different signal computation can help us better handle the systematics of the signal.
As shown in Fig,. \ref{fig:nuproduction}, our approach agrees well with WimpSim and Cirelli et al, while the spectrum obtained with PPPC differs considerably, mainly for the softer channel. This difference can be caused by the treatment of bottom and charm hadron interactions and the EW corrections. A full study of the origin of this discrepancy is underway.

\subsubsection{Neutrino Flux after Propagation}
\begin{figure}[!htb]
\centering
\includegraphics[width=0.28\linewidth]{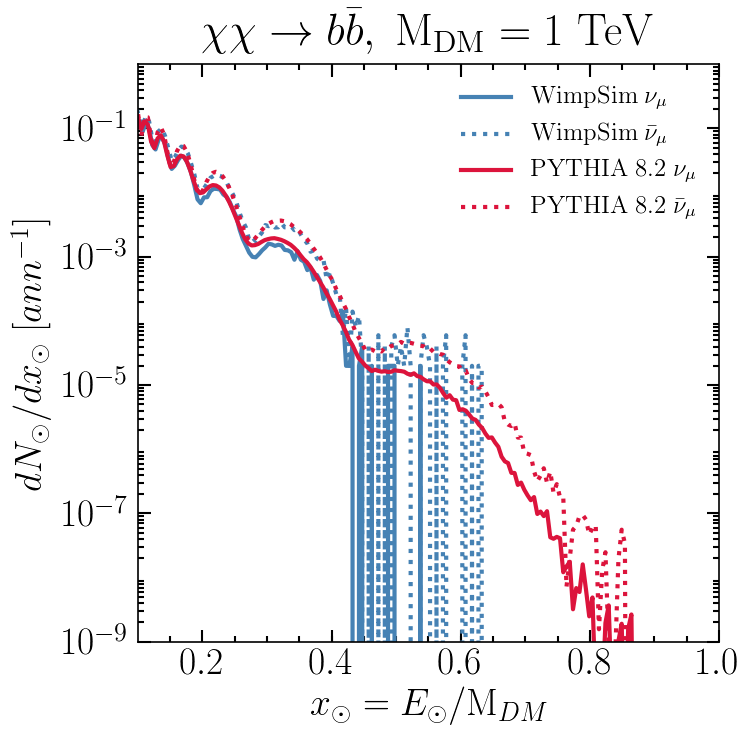}
\includegraphics[width=0.28\linewidth]{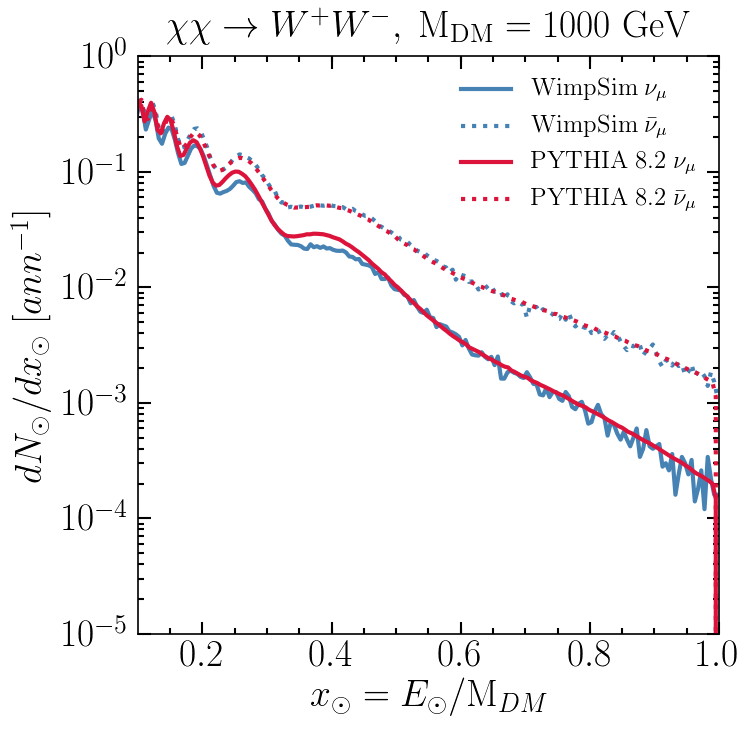}
\includegraphics[width=0.28\linewidth]{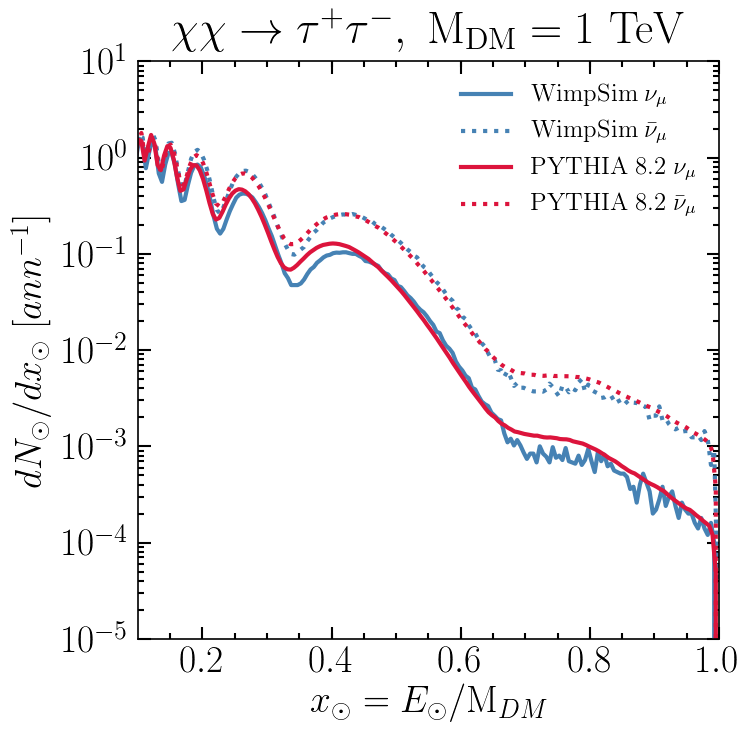}
\caption{Muon neutrino fluxes per annihilation at the Sun surface with normal mass ordering. Blue lines are generated by WimpSim and red lines are generated using PYTHIA8240 and $\nu$SQuIDS. Both generations have oscillation parameters from \cite{Esteban:2018azc}. Both generators have the same Monte Carlo sample size.}
\label{fig:nusunsfc}
\end{figure}

\begin{figure}
%\vspace*{-1pt}
\centering
\includegraphics[width=0.28\linewidth]{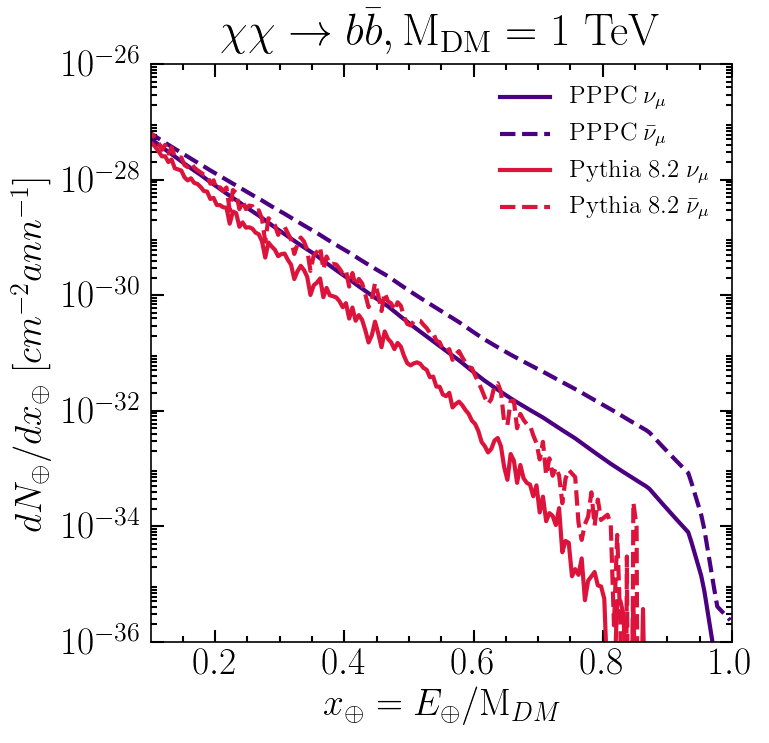}
\includegraphics[width=0.28\linewidth]{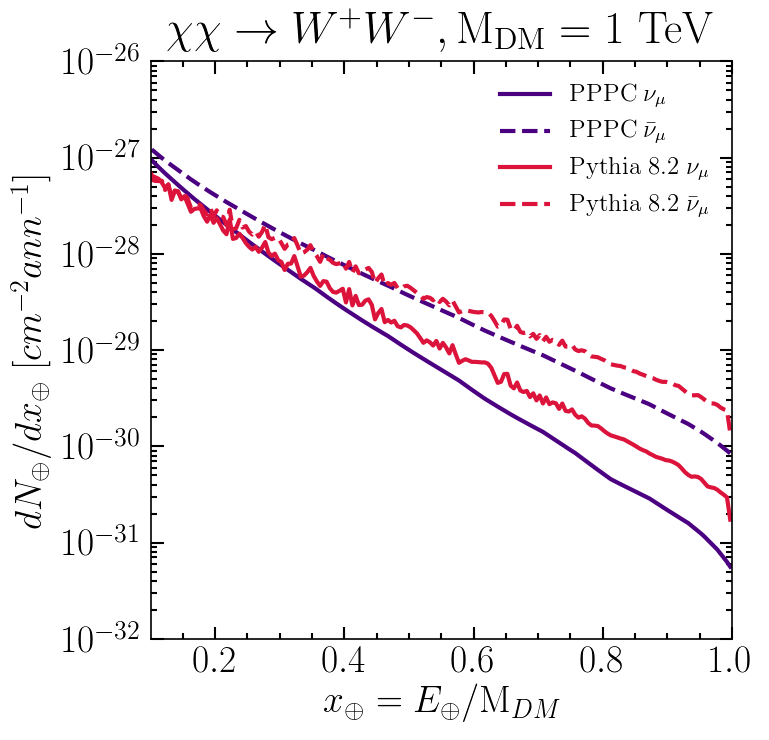}
\includegraphics[width=0.28\linewidth]{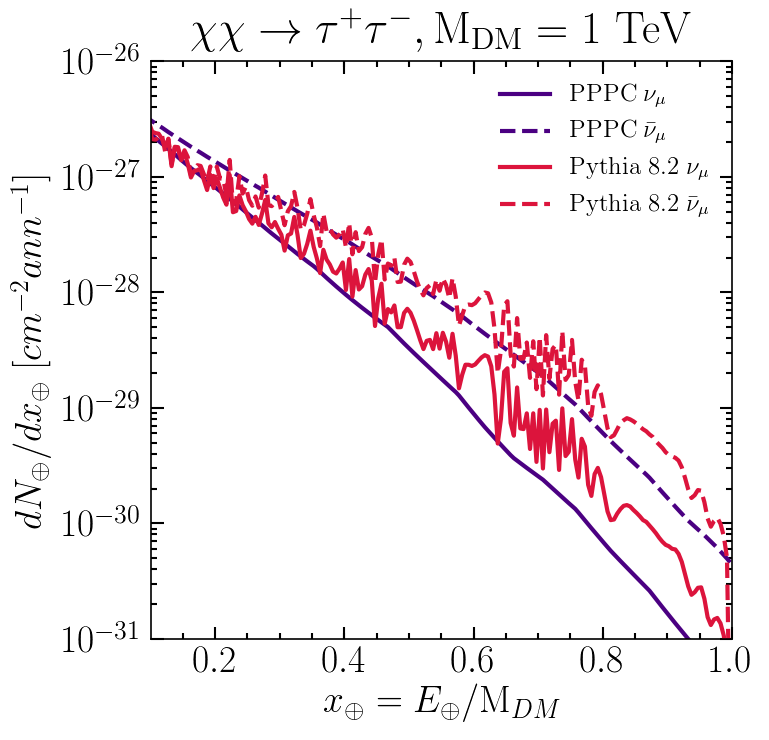}
\caption{Muon neutrino fluxes per annihilation per unit area at the location of IceCube with integrated zenith angles from $66.56^\circ$ to $113.44^\circ$. Indigo lines are from PPPC which used oscillation parameters from \cite{Capozzi:2013csa} and red lines are generated using PYTHIA8240 and $\nu$SQuIDS with oscillation parameters from~\cite{Esteban:2018azc} for normal mass ordering.}
\label{fig:nudetector}
\end{figure}

Neutrinos are produced in the solar core. They experience interactions and oscillations as they travel through the Sun, vacuum, and the Earth before reaching the detector. To study the oscillation effects, we use parameters reported by NuFIT~\cite{Esteban:2018azc} and compute neutrino propagation with a modified version of $\nu$SQuIDS~\cite{Delgado:2014kpa,arguelles:2015nu} that does not use the default iso-scalar neutrino cross section approximation to compare results with WimpSim and PPPC. From the core to the surface of the Sun, an isotropic spectrum is obtained, which is shown in Fig.~\ref{fig:nusunsfc}. We find good agreement in both neutrino and antineutrino propagated fluxes, with small differences for the harder channels at low energies. This could be a result of different implementation of tau-regeneration ($\nu$SQuIDS assumes polarized $\tau$ decays, WIMPSim unpolarized), or interpolation of the differential neutral-current neutrino cross sections. After neutrinos get through the Sun, surviving neutrinos continue traveling to the Earth. The flux of neutrinos at the detector depends on the detector location, {\textit i.e.} the zenith angle of the Sun. For IceCube at the geographic South Pole, the zenith angle of the Sun ranges from 66.56$^\circ$ to 113.44$^\circ$. Fig.~\ref{fig:nudetector} shows a comparison between the flux at the detector with integrated zenith angles of the Sun location per annihilation.

\subsection{Event Selection}
This analysis uses an event selection that has been optimized for TeV neutrinos, while maintaining a high purity of muon neutrinos and anti-neutrinos. This sample was developed for IceCube's latest sterile neutrino search and is described in detail in~\cite{Aartsen:2015rwa,TheIceCube:2016oqi}.

This event selection collects events with  >99.9\% purity of muon neutrinos at a rate of $\sim$ 1.4 mHz. 
For this purpose, the selection is restricted to the events from the Northern Sky. This greatly reduces the number of atmospheric muons because of the large overburden below the horizon.
Additionally, the selection requires that the number of DOMs triggered in an event be greater than a zenith dependent number, with DOMs at smaller zenith angles requiring more triggered DOMs, in order to remove misreconstructed events.
The resulting event selection covers an energy range from 200 GeV to 10 TeV in muon energy proxy and ranges in cosine of the zenith angle from -1 to 0.1.

In addition to greater statistics due to longer livetime, this event selection also includes an updated and improved treatment of systematics.
The latest models for both the bulk ice, as well the most recent DOM efficiencies are included.
Furthermore the updated "hole ice" model---ice that refroze after drilling cable holes for the DOMs with properties differing from the bulk ice---is also implemented.
Since three of the five major sources of error in the previous IceCube solar WIMP search were due to ice effects or DOM efficiency, we hope that this updated treatment of these properties will improve the capabilities of our analysis.
\subsection{Signal and background distributions}
\begin{wrapfigure}{r}{0.4\textwidth}
  \begin{center}
    \includegraphics[width=0.38\textwidth]{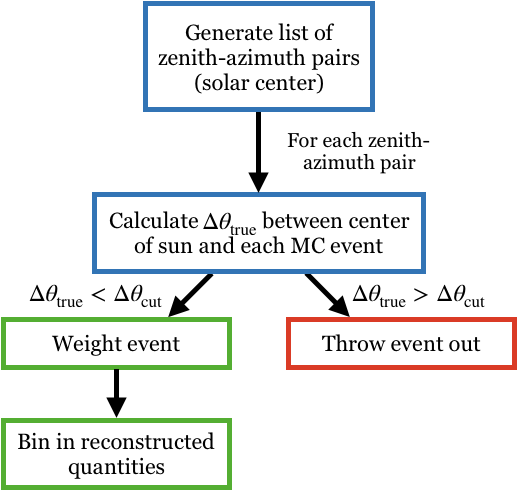}
  \end{center}
  \caption{Flow chart showing how signal distributions are generated. $\Delta\theta_{\rm{cut}}$ is the angular distance between the center and edge of the sun.}
  \label{fig:sigdistflowchart}
\end{wrapfigure}
We estimate the expected atmospheric neutrino background using MCEq \cite{Fedynitch:2015zma}. Here, we use the  Sybill2.3c \cite{Riehn:2017mfm} interaction model and Hillas-Gaisser-H3a (2012) \cite{Gaisser:2011cc} as the primary cosmic-ray model.

The dominant uncertainties can be parameterized by an overall normalization of the background flux, a change in the cosmic-ray spectral index, and uncertainties in the hadronic interaction models. For the latter we intend to use the Barr scheme \cite{Barr:2006it}.
In this scheme, different parts of the proton-air to meson differential cross section are modified by multiplicative factors.

\begin{figure}[h]
\centering
\captionsetup[subfigure]{justification=centering}
    \begin{subfigure}[t]{.38\textwidth}
      \centering
      \includegraphics[width=.9\linewidth]{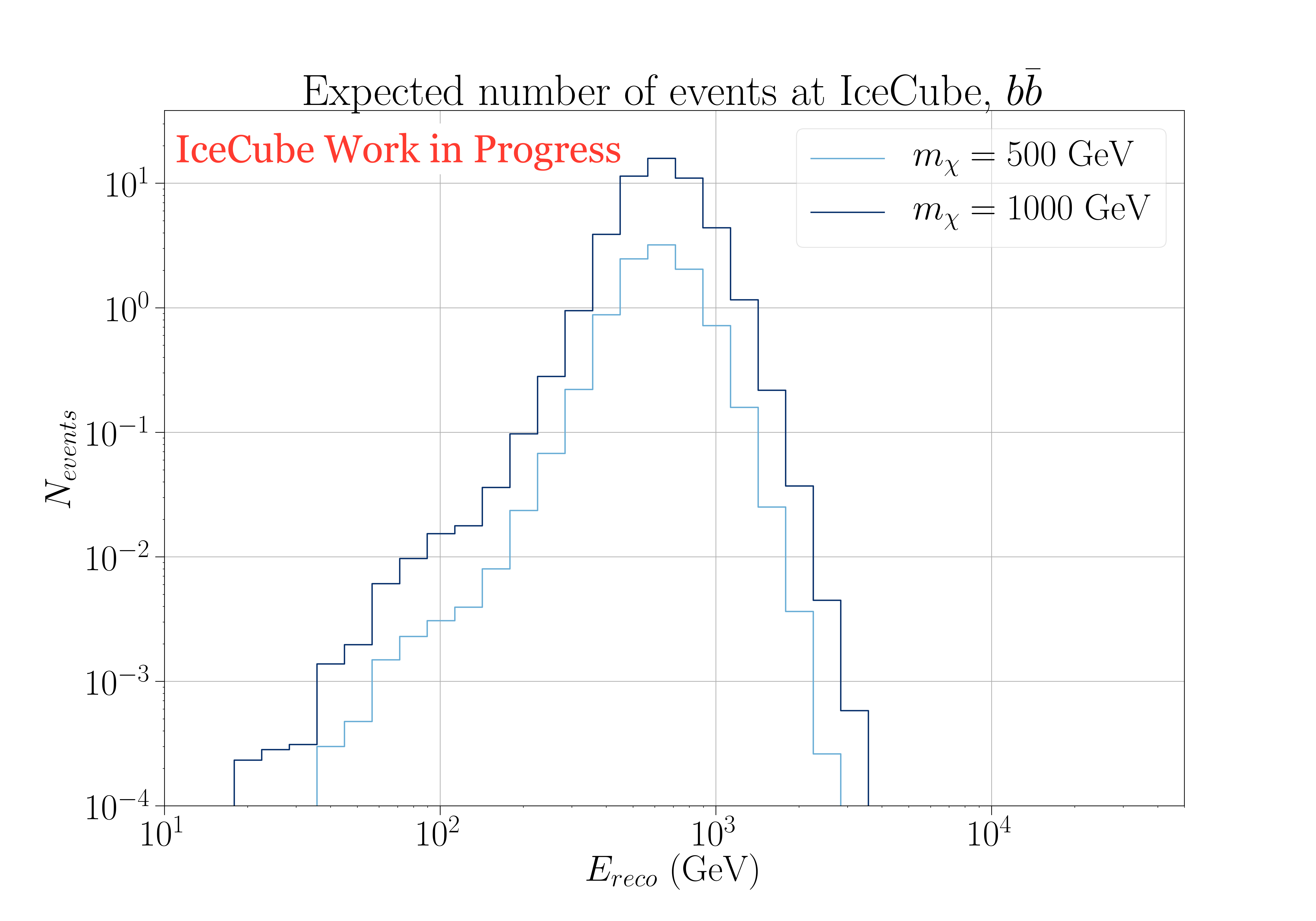}
      \caption[fontsize=small]{$\chi\chi\rightarrow b\bar{b}$,\\ $\sigma_{\chi \rm{N}, 500}^{\rm{SD}} = 3.06\times10^{-3}$ pb and \\ $\sigma_{\chi \rm{N}, 1000}^{\rm{SD}} = 2.59\times10^{-3}$ pb.}
    \end{subfigure}%
    \begin{subfigure}[t]{.4\textwidth}
      \centering
      \includegraphics[width=.9\linewidth]{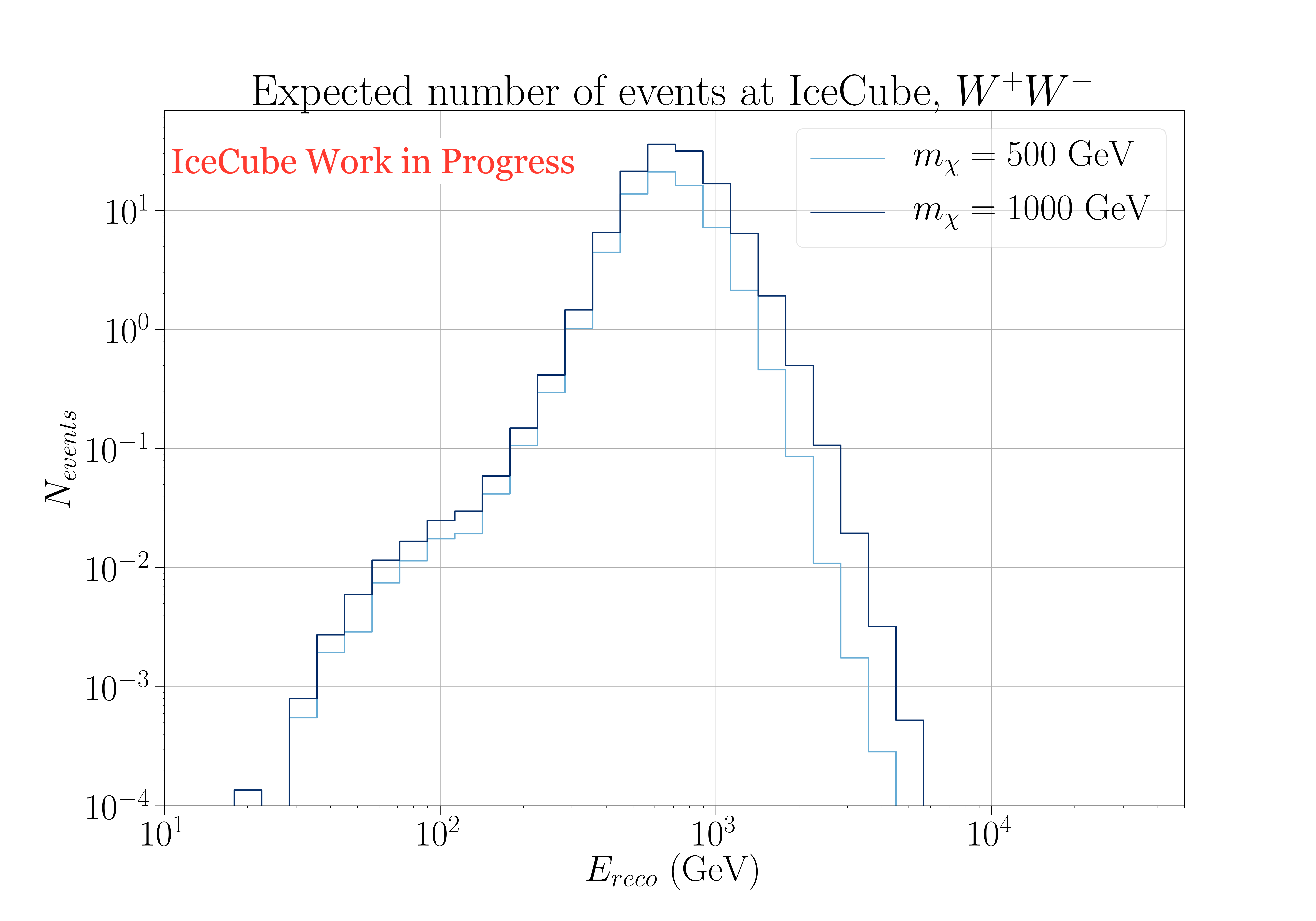}
      \caption[fontsize=small]{$\chi\chi\rightarrow W^{+}W^{-}$,\\
      $\sigma_{\chi N, 500}^{SD} = 3.76\times10^{-5}$ pb and \\
      $\sigma_{\chi N, 1000}^{SD} = 6.80\times10^{-5}$ pb.}
    \end{subfigure}
    \caption{Expected distribution of signal events from seven years of data taking.} 
    \label{fig:bgdists}
\end{figure}

\begin{figure}[h]
\centering
\captionsetup[subfigure]{justification=centering}
    \begin{subfigure}[t]{.38\textwidth}
      \centering
      \includegraphics[width=.9\linewidth]{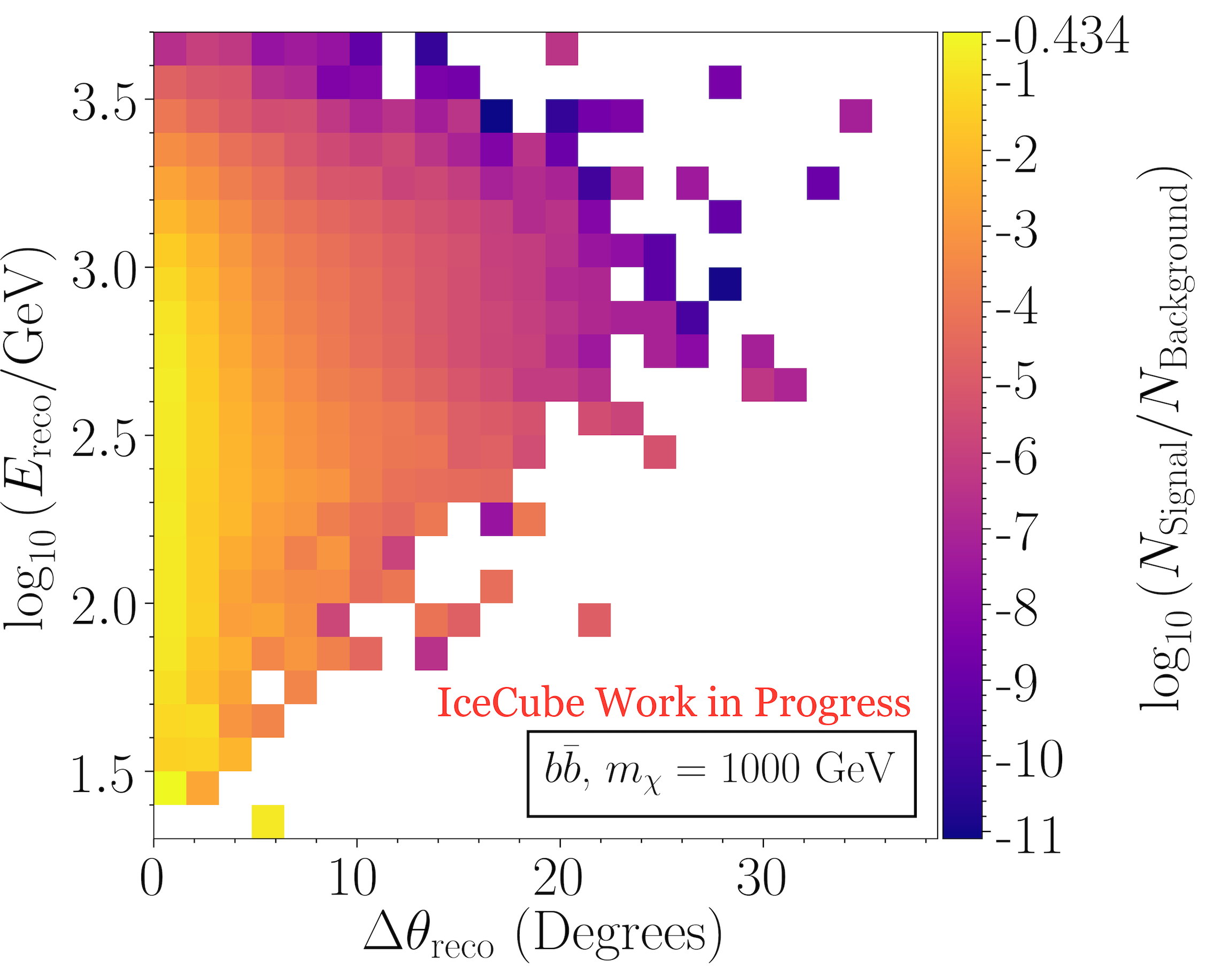}
      \caption[fontsize=small]{$\chi\chi\rightarrow b\bar{b}$, $m_{\chi}$=1000 GeV, and\\ $\sigma_{\chi N}^{SD} = 2.59\times10^{-3}$ pb.}
    \end{subfigure}%
    \begin{subfigure}[t]{.4\textwidth}
      \centering
      \includegraphics[width=.9\linewidth]{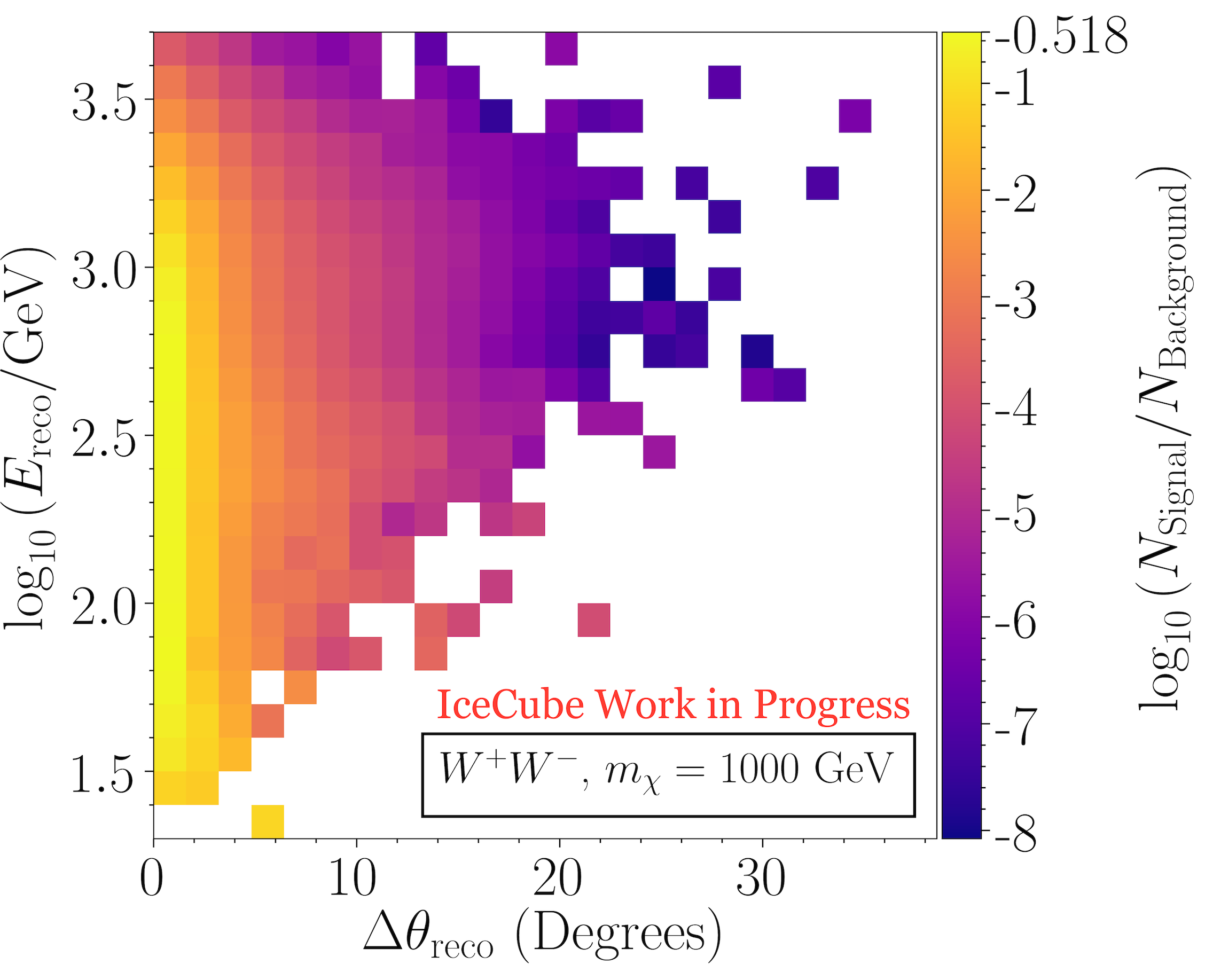}
      \caption[fontsize=small]{$\chi\chi\rightarrow W^{+}W^{-}$, $m_{\chi}$=1000 GeV, and $\sigma_{\chi N}^{SD} = 6.80\times10^{-5}$ pb.}
    \end{subfigure}
    \caption{Expected distribution of signal events divided by background events from seven years of data taking.}
    \label{fig:signaldists}
\end{figure}

The signal distributions are computed by first generating a large set of true azimuth-zenith pairs, where the azimuths are sampled uniformly from $0$ to $2\pi$, and the zeniths are weighted according to a distribution of the solar position.
The true angular distance from the center of the Sun, $\Delta\theta_{\rm{true}}$, is then computed for each pair.
If $\Delta\theta_{\rm{true}} < \Delta\theta_{\rm{cut}}$---where $\Delta\theta_{\rm{cut}}$ is the angular distance from the center to the edge of the Sun---the event is weighted by the expected neutrino flux from WIMP annihilations.
These are then binned in terms of reconstructed quantities. See Fig. \ref{fig:sigdistflowchart} for a flowchart of this process.
Examples of signal distributions as a function of $E_{\rm{reco}}$ and of signal divided by background as a function of $E_{\rm{reco}}$ and $\Delta\theta_{\rm{reco}}$ are shown in Figures \ref{fig:bgdists} and \ref{fig:signaldists}. The cross sections have been set to the current best limits from the previous IceCube solar WIMP search \cite{Aartsen:2016zhm}.
\section{Summary and Outlook}
In this work, we presented the theoretical underpinnings of indirect detection searches for solar WIMPs in IceCube as well as provided a brief description of the previous IceCube search and its limiting factors.
We then discussed our analysis chain and highlighted the main ways in which our current solar WIMP search intends to improve upon the previous one.
Specifically, we have performed an in-depth comparison of different event generators in order to understand the impact of our choice on the systematic uncertainty.
In addition to this study, we use an event selection that has more than twice the livetime and has incorporated the most up to date systematic treatment.
Furthermore, we intend to study the effect of solar atmospheric neutrinos in order to more realistically characterize our background.

\section*{Acknowledgements} \label{sec:ack}

We would like to thank Joakim Edsj\"o for useful and engaging discussions during the conference.

\bibliographystyle{ICRC}
\bibliography{references}

\end{document}